\documentstyle[a4,times,epsf]{article}

\title{The breakdown flash of Silicon Avalance Photodiodes -- backdoor for
  eavesdropper attacks?}
\author{Christian Kurtsiefer$^1$,  Patrick
  Zarda$^2$, Sonja Mayer$^1$, and Harald
  Weinfurter$^{1,2}$\\[2mm] {\small\it
$^1$Sektion Physik, Ludwig-Maximilians-Universit\"at, D-80799 
  M\"{u}nchen, Germany}\\[1mm]
{\small\it $^2$ Max-Planck-Insitut f\"{u}r Quantenoptik, D-85748 Garching,
  Germany}
}

\date{\today\\[3mm] {\small\it submitted to J. Mod. Opt.}}

\begin{document}
\maketitle

\begin{abstract}
Silicon avalanche photodiodes are the most sensitive photodetectors in the
visible to near infrared region. However, when they are 
used for single photon detection in a Geiger mode, they are known to emit
light on the controlled breakdown used to detect a photoelectron. This
fluorescence light might have serious impacts on experimental applications
like quantum cryptography or single-particle spectroscopy. 
We characterized the fluorescence behaviour of silicon avalanche photodiodes in
the experimentally simple passive quenching configuration and discuss
implications for their use in quantum cryptography systems.
\end{abstract}

% \pacs{}
\section{Introduction}
For a long time, silicon avalanche photodiodes (APD) have been used for single
photon detection in the near-infrared region\cite{conrad68,haecker71} because
of their high quantum efficiency and low dark count rate. These properties are
particularly important for quantum cryptography\cite{lo99,bou00,crypt1,zbi98},
where a huge yield of secure bits and a low signal/noise ratio is crucial.

To obtain a single photon counting behaviour, the avalanche diode is operated in
an all-or-nothing counting mode similar to the way Geiger detectors are used
in nuclear physics for particle counting. In this so-called Geiger mode, the
diode is reversely biased
above the breakdown voltage such that a single photoelectron can generate a
self-sustaining discharge. The discharge current is used as an
indicator for the generation of a photoelectron and thus of an absorbed
photon. Thereby, a timing accuracy better then 60~ps has been
achieved\cite{cova81}.

It has been observed previously that the avalanche of
charge carriers is accompanied by photon emission\cite{raritypriv}. Although
this light
emission is not very strong, in several single photon counting applications it
may have serious impacts on the experiment. In quantum cryptography, for
example, such a light emission might enable an external observer to gain
information of a photo detection event on the receiver side, opening a possible
eavesdropping back door to an otherwise secure communication channel.
Another experimental situation in which this photoemission has to be
considered are photon correlation measurements, as they are
performed in single atom or molecule spectroscopy. In a typical
Hanburry-Brown--Twiss configuration, two photodetectors 
are looking onto a faint light source, and one has to ensure that light
emitted in the breakdown flash of one photodiode is not causing artificial
photo events in the second photodetector due to residual crosstalk between the
two photodetectors\cite{diam,orrit,lounis}. It is therefore important to know the
photoemission characteristics of that breakdown photoemission to avoid
crosstalk with the light to be detected. In this paper, we describe our
investigation of the temporal and spectral
distribution as well as the absolute amount of light emitted during a
detection event.

\section{Photodiode operation}
A photodetection process is initiated by a photoelectron created after
absorption of a photon in a reverse-biased $pn$-junction. This electron is
accelerated into a highly doped region where an avalanche of charge carriers
is triggered. In single photon counting mode, the bias voltage exceeds the
breakdown voltage of the diode, 
meaning that once an avalanche has been triggered, it is self-sustaining as
long as the external voltage exceeds the breakdown threshold. To avoid the
thermal damage of the diode and to bring it back into a state ready 
for a subsequent photoelectron detection, the avalanche has to be
quenched. This is done by lowering the reverse bias voltage
across the diode for a certain time. After allowing all charge carriers to
recombine and thus bringing the diode into an insulating state again, a full
photodetection cycle is finished and the diode is ready for the next
event. 

\begin{figure}
\centerline{\epsfxsize11cm\epsffile{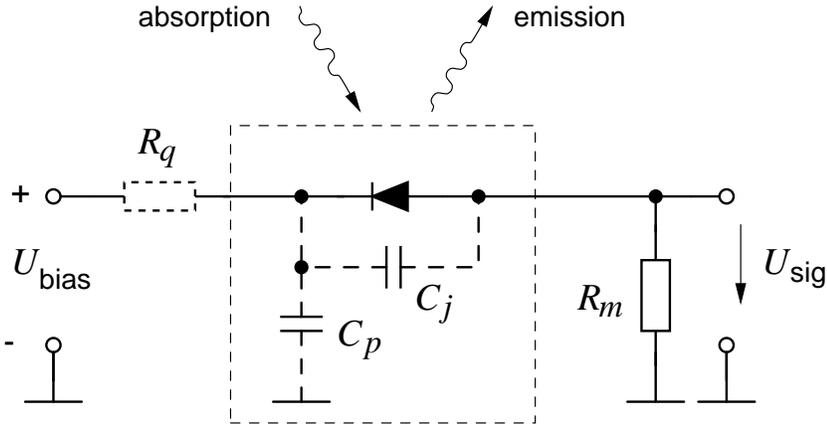}}
\caption{\label{diodesetup}
Operation of the APD in passive quenching mode. The diode (with a junction
capacity $C_j$ and a parasitic capacity $C_p$) is reverse biased via a
high impedance quenching network $R_q$ above the breakdown voltage. At diode
breakdown, the parasitic capacity $C_p$ discharges through the load
resistance $R_m$, causing a voltage peak indicating the breakdown.
}
\end{figure}

The usual
configurations for that procedure are referred to passive and active
quen\-ching\cite{brown87}. In passive quenching, the diode is reverse-biased via
a low-current network (e.g. a large resistor $R_q$) such that the discharge
current triggered by a photoelectron avalanche causes a voltage drop, reducing
the voltage across the diode below the breakdown
voltage (see figure \ref{diodesetup}). Then, the 
junction capacity $C_j$ has to be  recharged again to the full reverse bias
voltage. With usual passive quenching configurations, a recharge time on the
order of a microsecond is achieved. To obtain a faster recharge and thereby a
shorter dead time, active quenching techniques have to be used\cite{cova81}.
Yet, the discharge current and thus the breakdown flash should not depend on
the quenching configuration.

% $<<${\bf figure 1 about here }$>>$

In our experiments, we used an APD with an integrated two stage thermoelectric
cooler, type C30902-SDTC from Perkin-Elmer. The diodes have a circular active
area of 0.5~mm 
diameter, and are accessible through a transparent window. They are mounted in
modules together with a high voltage supply, a discriminator to generate
standard NIM pulses and a temperature controller for the peltier
element\cite{patrickthesis}. We use a current limiting network instead of a
quenching resistor, a measurement resistor of $R_m=100\;\Omega$ and a reverse
bias voltage of $U_{\rm bias}=215$~V at a temperature of $-25~^\circ$C, which
is approximately 20~V above the breakdown of the APD. According to the
manufacturer, the diodes 
are supposed to show a single photon detection efficiency of up to $55$\% at a
wavelength of 800~nm\cite{eggdata}, depending on operating conditions.

% $<<${\bf figure 2 about here }$>>$

\begin{figure}
\centerline{\epsfxsize11cm\epsffile{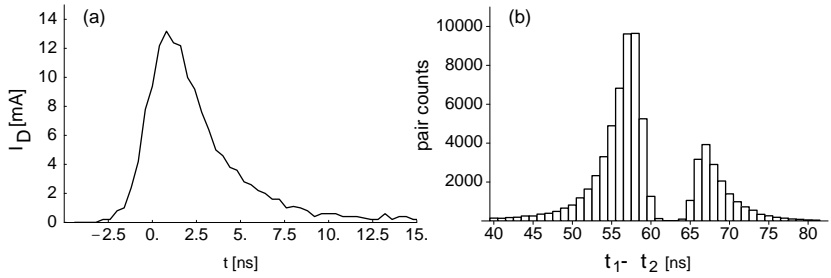}}
\caption{\label{histofit}
(a) Discharge current $I_D(T)$ of the avalanche photodiode during a breakdown
cycle. (b) Histogram of coincidence events of both photodetectors. The left peak
corresponds to photons emitted by diode 1 seen by diode 2. Both peaks have an
exponential decay with a time constant of 2.9~ns. The asymmetry is due to the
different magnification of the two diodes looking at each other.
The shape of each peak resembles very much the discharge current behaviour.
}
\end{figure}

The discharge current $I_D(t)$ we measured under these conditions is shown in
figure~\ref{histofit}a. It reflects an exponential decay, convoluted with a
Gaussian distribution. From this measurement, we obtain a total charge of 
$$Q_D=\int T_D(t)\,dt=64\;{\rm pC}$$
released during a diode breakdown. From that value, we deduce a  parasitic
capacity $C_p$ of
$$C_p=Q_D/20~{\rm V}=3.2~{\rm pF}\quad,$$
assuming that during breakdown, most of the current through $R_m$ is supplied
by $C_p$, and not by the biasing network.

\section{Absolute photoemission rate}
\begin{figure}
\centerline{\epsfxsize11cm\epsffile{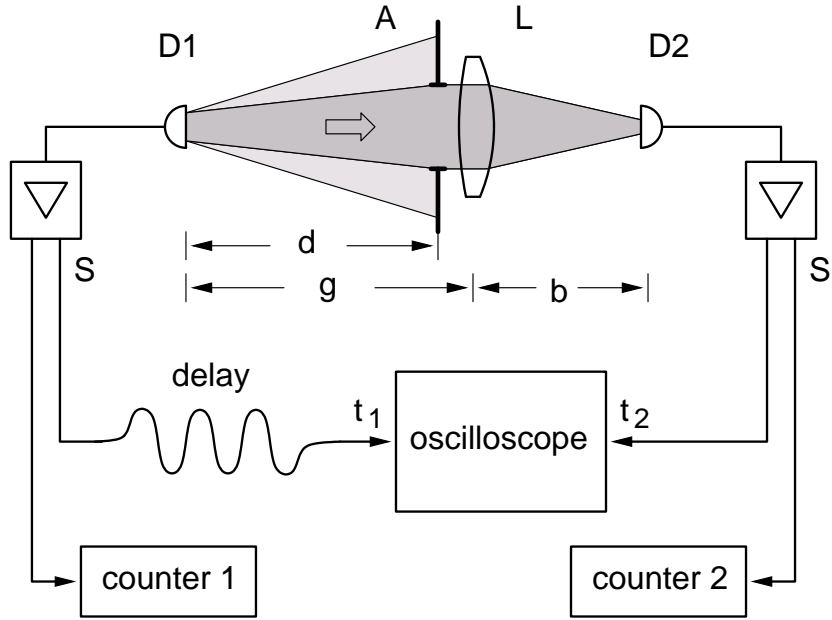}}
\caption{\label{mutual}
Setup to determine the integral fluorescence light. The active area of
photodiode $D_1$ under 
investigation is imaged through a lens $L$ onto a second photodiode $D_2$,
with an aperture $A$ defining the solid angle seen of diode $D_1$. Both diodes
are operating in passive quenching Geiger mode. Their breakdown current is
amplified, discriminated and sent to counters and an oscilloscope to
investigate coincident events (The signal of $D_1$ is delayed by
63~ns).
}
\end{figure}

To determine the amount of light emitted during a breakdown cycle, we used an
optical arrangement sketched in figure~\ref{mutual}. The active area of an APD
module $D_1$ is imaged with a lens $f=50$~mm onto a second APD, $D_2$, with a
demagnification of 2 (corresponding to distances of $g=150$~mm and $b=75$~mm,
respectively). This ensured that light emitted from all parts of the active
area of diode $D_1$ could reach the active area of $D_2$ even for imperfect
alignment. To define the solid angle of light collected from diode $D_1$, we
used an aperture $A$ with a diameter of 3~mm (and 5~mm in a second experiment)
at a distance of $d=123$~mm from the diode. The corresponding solid angles are
$\Omega_3=4.67\cdot10^{-4}$~sr and $\Omega_5=1.3\cdot10^{-3}$~sr, respectively.

% $<<${\bf figure 3 about here }$>>$

The NIM pulses were sent both to PC-card based
counters, and for timing analysis to a digital oscilloscope (LeCroy
LC574A). Using a pair trigger feature together with an auxiliary delay line
of 63~ns, we collected coincidence events of the two detectors and
histogrammed their time differences $t_1-t_2$ in an interval of -40~ns to
+60~ns with a resolution of a few 100~ps.

To measure only the light emitted by the diode $D_1$,  we lowered the
ambient light such that $D_1$ registered a count rate of $r_1=2634$~cps. This
is only moderately larger than the dark count rate (approximately 500~cps) and 
ensures that scattering of external light to the second diode $D_2$ is
minimal. With the optical path open to the second diode and an aperture
diameter of 3~mm, we observe a count rate from diode $D_2$ of $r_2=731$~cps.
Finally, only pair events are selected which ensures that only the properties
of the breakdown flash were analysed.

A histogram of time differences for double photo events 
is shown 
in figure~\ref{histofit}b. One can clearly recognize two peaks, the left one
($t_1-t_2<60$~ns) corresponding to photo events registered in detector $D_2$
after a discharge of detector $D_1$, and the right one corresponding to
the reverse process. The
asymmetry in the amplitudes of the two peaks can be explained by the
asymmetry in the imaging optics, as the aperture is not located exactly at the
lens position, and/or by a difference in the amount light produced by the two
diodes. Each peak shows a rise time between 1 and 2 ns, and an exponential
decay, probably following the discharge current of the diode.
The distribution $h(\Delta t)$ of each peak of time differences $\Delta
t=t_1-t_2$ in figure~\ref{histofit}b 
has the same temporal pattern as the discharge current shown in
figure~\ref{histofit}a. 

We modeled a higher resolution histogram of the first
peak of $h(\Delta t)$ by a convolution of an
exponential decay with a time constant $\tau$ 
and a Gaussian distribution with a variance $\sigma$. Using the model function
$$h(\Delta t)= \left(\Theta(\Delta t)\cdot e^{-\Delta
  t/\tau}\right)\otimes\left(e^{-\Delta t^2/(2\sigma^2)}\right)\quad,$$
where $\Theta(t)$ is a step function, we obtain fit values of
$\tau=2.75\pm0.07$~ns and $\sigma=0.72\pm0.03$~ns. The actual shape of this
distribution is determined by the discharge network.

Integration over the pair distribution from $t_1-t_2=20$~ns to 62~ns leads to a
rate of $n_c=48.4\pm1\;{\rm s}^{-1}$ for photo events of detector $D_2$
induced by breakdown events of $D_1$; the accidental count rate for that time
window, 
$$n_{acc}=r_1r_2\cdot42~{\rm ns}=0.081~{\rm s}^{-1}$$
 is negligible. With  the breakdown rate $r_1$ of $D_1$
and the captured solid angle $\Omega_3$, and assuming isotropic emission of
the fluorescence light, we obtain a differential breakdown emission intensity
of 
$${d\,n_L\over d\,\Omega}={n_c/r_1\over\Omega_3}=39\;{\rm photons/sr}$$
for each detected breakdown of the diode $D_1$.
From a similar measurement with an aperture diameter of 5~mm, we found a value
of $dn_L/d\Omega=43$ photons/sr. Within the accuracy of the alignment of the
photodetectors and the assumption of isotropy of emission, these two values
are compatible. However, these values do not contain a detection efficiency
$\eta$ yet. Because this detection efficiency varies with the wavelength (and
has a maximum of $\approx55\%$ at $\lambda\approx820$~nm\cite{eggdata}), an
estimate of total rate can only be given with a knowledge of the spectral
distribution.

\section{Spectral distribution of the breakdown emission}
\begin{figure}
\centerline{\epsfxsize11cm\epsffile{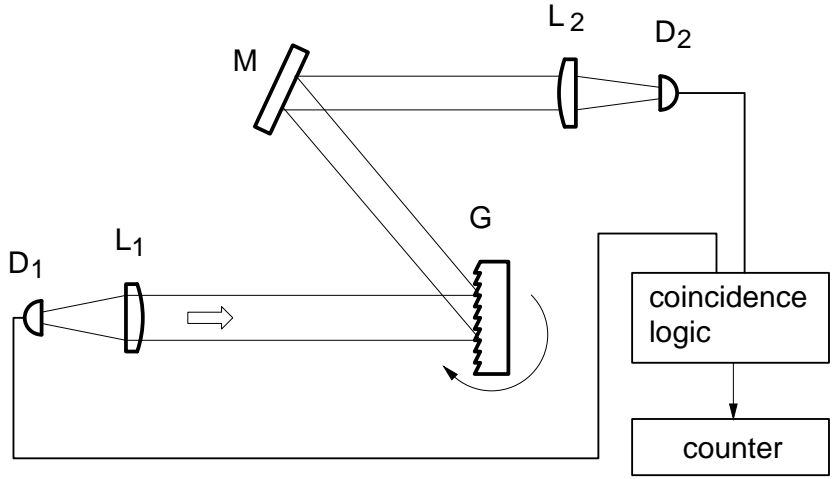}}
\caption{\label{spektrometer}
Setup for measuring the breakdown flash spectrum. Light emitted form a diode
$D_1$ is collimated through a lens $L_1$ with a focal length of $f=150$~mm,
dispersed through a diffraction grating $G$ and detected after a fold mirror
$M$ with a second single photon counting APD $D_2$ after a collimation lens
$L_2$. The spectrum becomes visible in coincidence events between the two
detectors $D_1$ and $D_2$.
}
\end{figure}

In order to evaluate possible countermeasures in experiments sensitive to the
breakdown light emission of an APD, we measured the spectral
distribution of that light. Therefore, we used again a setup of two single
photon counting APDs looking at each other, where we inserted a reflection
grating as a tunable filter in the optical path as shown in
figure~\ref{spektrometer}. The active area of the diode under investigation,
$D_1$, was placed in the focal plane of a lens $L_1$ ($f=150$~mm) to collimate
the light emitted in a diode breakdown. The first diffraction order of a
blazed grating (1200 lines/mm) was focused with another lens $L_2$ onto the
second APD, $D_2$, acting as a photon detector. At a wavelength of 632~nm, we
thereby obtain a wavelength resolution of approximately 3.3~nm FWHM; we
adjusted the transmitted wavelength by turning the grating.

% $<<${\bf figure 4 about here }$>>$

Again, we identify photons from the breakdown flash in $D_1$ by looking for
coincidences of detector events in $D_1$ and $D_2$. We have chosen a
coincidence time window of $\tau_c=70$~ns after a breakdown of $D_1$. In 
the experiment, we recorded the number of coincidence events, $N_c(\lambda)$,
and events $N_1(\lambda),N_2(\lambda)$ of the individual detectors for an
integration time $T$. To
obtain acceptable signal levels, we exposed detector $D_1$ to a raised level
of background light, causing breakdown rates of
$N_1/T\approx17000\ldots20000~{\rm s}^{-1}$. The corresponding count rate
$N_2/T$ of detector $D_2$ was in the range of $5000\ldots6000~{\rm
  s}^{-1}$. The number of coincidence events varied form 300 to 1100 counts
over the recording periode. 

\begin{figure}
\centerline{\epsfxsize11cm\epsffile{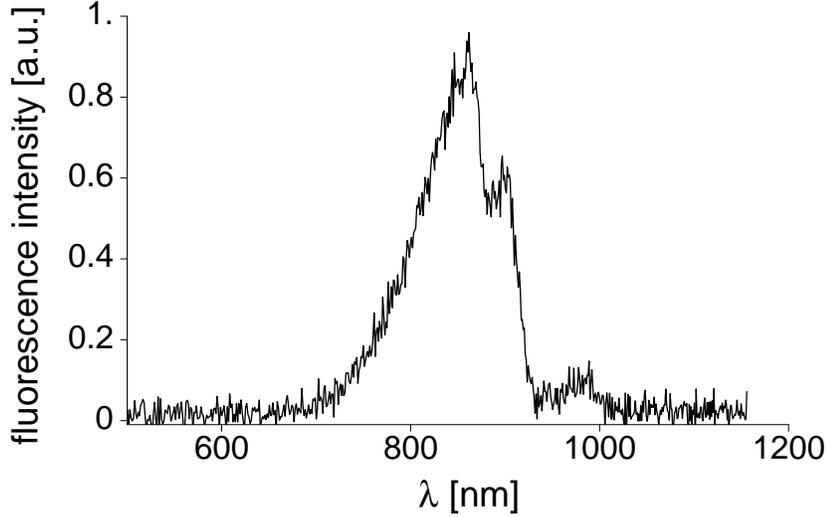}}
\caption{\label{spektrum}
Normalized breakdown flash spectrum of the silicon avalanche photodiode. The
emission is peaked around 850~nm. The curve shown is obtained by subtracting
the accidental coincidence rate from the raw measured coincidences and
subsequent normalization to the count rate of the emitting diode.
}
\end{figure}

We correct for accidental coincidences and fluctuations in the breakdown rates
of APD $D_1$, and obtain a normalized spectral distribution $I(\lambda)$ from
our experimental data using the expression:
$$I(\lambda)=\alpha{N_c(\lambda)-N_1(\lambda)\cdot N_2(\lambda)\cdot
  \tau_c/T\over N_1(\lambda)}$$
The spectrum obtained after an integration time of $T=50~$sec per point is
shown in figure~\ref{spektrum}, using a normalization constant of
$\alpha=10^{3}$.

% $<<${\bf figure 5 about here }$>>$
 
One can identify a spectral emission ranging from 700~nm to 1000~nm, with a
maximum at 860~nm, two sharp edges at 872~nm and 913~nm, respectively, and two
weaker maxima at 900~nm and 980~nm, respectively. 
This structure is a product of the emission
spectrum of the breakdown light, the transfer function of our spectrometer
setup and the spectral sensitivity for photo detection of the second avalanche
diode $D_2$.
While the transmission of the spectrometer is reasonably flat over the
investigated region, the main deviation between the measured and the emitted
spectrum can be
attributed to the wavelength dependency of the quantum efficiency
$\eta(\lambda)$ of detector $D_2$, which, according to the manufacturer,
has a smooth drop-off from 70\% to 8\% in the range of $\lambda=800$ to
1000~nm\cite{eggdata}. However, the key structures of the spectrum obtained
are not an artefact of the detection efficiency, and are
characteristic to the generation process of the emitted light.

\section{Impact of photoemission on a quantum cryptography system}
\begin{figure}
\centerline{\epsfxsize11cm\epsffile{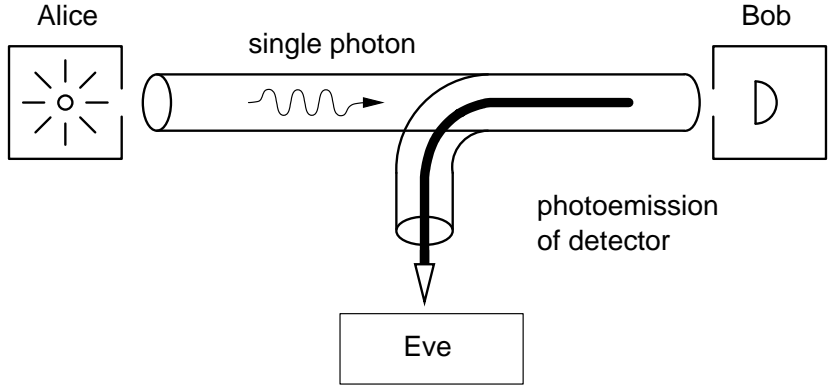}}
\caption{\label{eveattack}
Possible eavesdropping attack to quantum cryptography system. A single photon
carrying information on phase or polarization is sent from Alice to Bob
without interception by Eve. However, Eve could have access to the photons
emitted by Bob upon detection, and gain timing and/or polarization information
of the detected photon.
}
\end{figure}

In our experiments, we tried to quantify the photoemission on breakdown of
silicon avalanche photodiodes in Geiger mode. This photoemission may allow a
possible eavesdropper in a quantum cryptography application to gain
information of the outcome of a measurement simply by looking at this
photoemission, as sketched in figure~\ref{eveattack}.
It therefore has to be ensured that the amount of light leaking back to a
possible eavesdropper is small in order to limit its knowledge on the outcome
of the single particle measurement by Bob.

% $<<${\bf figure 6 about here }$>>$

To minimize the amount of light generated in the first place, the
capacity $C_P$ should be reduced to a minimum. This technique, however,
quickly reaches a limit with currently available photodiode packages.

Another measure to reduce the emitted light would be the use of optical
filters,
blocking the spectral range of 700-1000~mn in which photoemission
occurs. However, this technique is restricted to cases where the wavelength
of the transmitted light is outside that range. This is the case with recently
developed diamond-based single photon sources\cite{diam}, or using shorter
wavelength laser diode emission\cite{LANL}. For
systems using laser diodes around 850~nm exploring an absorption minimum in
optical fibers\cite{zarda2}, this technique would require narrow band
interference filters around the emission wavelength of the diodes. Then, the
possible leakage of information to an eavesdropper can be made negligibly
small, too.

Additionally, spatial filtering may be used to block light propagating back
the quantum channel. Assuming that the photoemission light is
emitted without spatial coherence across the photo detection surface, and that
light to be detected is coming out of a single spatial mode from an optical
fiber or an equivalent spatial mode filter in a free space arrangement, the
back-propagating light is reduced.

To estimate the fraction of light coupled back, we first consider the
breakdown flash 
brilliance (i.e., the number of photons emitted per surface area and solid
angle) for each photon detection event. From our measurements, we find 
$$B={dn_L\over d\Omega}\cdot{1\over A_{D}}=2\cdot10^{-3}\,{{\rm  photons}\over
  {\rm sr}\cdot \mu{\rm m}^2}\quad, $$
where $A_D$ is the sensitive area of the photodiode. 
 The number of photons $N_r$ collected from such an incoherent source into a
single spatial mode, characterized e.g. by a Gaussian beam waist $w_0$ and a
corresponding divergence $\theta_D$, is given by:
$$N_r=B\cdot\int\limits_{r=0}^{\infty}e^{-2r^2/w_0^2}\,rdr\cdot\int\limits_{2\pi}e^{-2\theta^2/\theta_D^2}\,d\Omega\approx
B\cdot w_0^2\Theta_D^2{\pi^2\over4}=B\cdot{\lambda^2\over4}$$
Integrating over a wavelength range from 700~nm to 1050~nm, we obtain 
a numerical value of $N_r=3.6\cdot10^{-4}$~photons coupled into the single
spatial mode of the quantum channel for a detection event. This value is
independent of the detailed structure of the coupling optics as long as
reciprocal optical elements are used. It is also only a lower limit obtainable
with a similar photodetector, since we have not taken into account the quantum
efficiency of the photo detector. 

To correct for the quantum efficiency and to estimate the real number of
photons coupled back into the quantum channel, we use the measured spectral
distribution
$I(\lambda)$ and a detection efficiency $\eta(\lambda)$ (i.e., the product of
photoelectron generation probability given by the manufacturer and the
photoelectron detection efficiency of .55 at 20~V above breakdown) obtained
from the manufacturer. Then, we numerically derive a correction factor given
by the expression:
$$\beta=
{\int\limits{I(\lambda)\over\eta(\lambda)}\,d\lambda\quad\left/\quad\int\limits
I(\lambda)d\lambda\right.}$$ 
For a wavelength range from 700~nm to 1050~nm, we obtain a
numerical value of $\beta\approx3.5$. With this factor, we end up with a
corrected numerical value of $N_r^{corr}=\beta N_r=1.3\cdot10^{-3}$~photons coupled back
into the single spatial mode of a quantum channel.

\section{Summary}
To summarize, we quantified the photoemission behavior of a silicon
avalanche diode during a breakdown, such as induced by a detection event of a
single photon, we found an emission spectrum ranging from 700~nm to 1000~nm,
and estimated the possible leak of information to a possible eavesdropper due
to this effect. Whereas this emission might have to be considered for single
atom and molecule spectroscopy, in quantum cryptography the backdoor for an
eavesdropper can be closed by taking some care with spectral and spatial mode
filtering. 
 It remains to be investigated if photodiodes used for quantum
cryptography systems\cite{gisin97} in the telecom wavelength range (1300~nm and
1550~nm) which are usually based on InGaAs or Ge, show a similar
effect. With InGaAs being a direct semiconductor, one could expect it to be
more likely for charge carriers to undergo radiative recombinations than in
silicon or germanium, thus showing a stronger breakdown flash.

%\acknowledgements
\section*{Acknowledgements}
This work was supported by the European Union in the EQCSPOT project (EC28139)
and the Deutsche Forschungsgemeinschaft.
 
%\begin{references}

\end{document}